\documentstyle[twoside,12pt,epsfig]{article}

\normalsize
\def\greaterthansquiggle{\raise.3ex\hbox{$>$\kern-.75em\lower1ex\hbox{$\sim$}}}
\def\lessthansquiggle{\raise.3ex\hbox{$<$\kern-.75em\lower1ex\hbox{$\sim$}}}
\newcommand{\beq}{\begin{equation}}
\newcommand{\eeq}{\end{equation}}
\newcommand{\beqa}{\begin{eqnarray}}
\newcommand{\eeqa}{\end{eqnarray}}
\newcommand{\ba}{\begin{array}}
\newcommand{\ea}{\end{array}}
\newcommand{\bce}{\begin{center}}
\newcommand{\ece}{\end{center}}

\newcommand{\grts}{\greaterthansquiggle}
\newcommand{\lets}{\lessthansquiggle}

\newcommand{\ra}{\rightarrow}

\def\t               {\theta}

\def\x               {\chi}
\def\ti              {\tilde}

\def\st              {{\ti t}}

\def\sb              {{\ti b}}

\def\sf              {\ti f}

\def\chp             {\ti \x^+}
\def\chm             {\ti \x^-}
\def\nt              {\ti \x^0}
\def\sg              {{\ti g}}
\newcommand\tht{{\theta_\st}}

\def\sf              {\ti f}

\def\chp             {\ti \x^+}
\def\chm             {\ti \x^-}
\def\nt              {\ti \x^0}
\def\sg              {{\ti g}}

\def\sth             {\sin\t}
\def\cth             {\cos\t}

\def\onehf           {{\textstyle \frac{1}{2}}}

\newcommand{\eq}[1]  {\mbox{(\ref{eq:#1})}}

\begin{document}

\begin{flushright}
UWThPh-1997-29\\
HEPHY-PUB 673/97\\
DESY 97-169\\
hep-ph/9709252\\
\end{flushright}
\vspace*{0.8cm}
\begin{center}

{\Large \bf  Production and Decay of Stops and Sbottoms,\\
  and Determination of SUSY Parameters\footnote{Contribution to the proceedings
  of  the ''ECFA/DESY Study on Physics and Detectors for the Linear Collider'', 
  DESY 97-123E, ed. R. Settles}
}

\vspace{1cm}

\begin{large}
  A. Bartl\footnote{bartl@Pap.UniVie.ac.at,
    $\odot$ helmut@hephy.oeaw.ac.at,
    $\star$ garfield@hephy.oeaw.ac.at,\\
    $\ddagger$ kraml@hephy.oeaw.ac.at,
    $\flat$ majer@hephy.oeaw.ac.at,
    $\natural$ porod@Pap.UniVie.ac.at,\\
    $\sharp$ andre.sopczak@cern.ch; \hspace{5mm}
    http://wwwhephy.oeaw.ac.at/p3w/theory/susy/}$^{\small 1}$,
  H. Eberl$^{\small \odot 2}$,
  T. Gajdosik$^{\small \star 2}$,
  S. Kraml$^{\small \ddagger 2}$,\\[1mm]
  W. Majerotto$^{\small \flat 2}$,
  W. Porod$^{\small \natural 1}$,
  A. Sopczak$^{\small \sharp 3}$ \\[8mm]
\end{large}

{\em (1) Institut f\"ur Theoretische Physik, Universit\"at Wien,
           A-1090 Vienna, Austria } \\[1mm]
{\em (2) Institut f\"ur Hochenergiephysik, \"Osterreichische Akademie der
           Wissenschaften, A-1050 Vienna, Austria } \\[1mm]
{\em (3) DESY-Zeuthen, D-15738 Zeuthen, Germany} \\

\end{center}

\vspace{8mm}

\begin{abstract}
We present numerical predictions for the decay branching ratios of the heavier
stop and sbottom mass eigenstates. We estimate the precision to be
expected for the determination of the underlying supersymmetry parameters
of the stop and sbottom systems.
\end{abstract}

\section{Introduction}
The study of pair production of scalar top quarks is particularly interesting
because the lighter stop $\st_1$ is expected to be the lightest 
scalar quark. 
Left--right mixing plays a r\^ole 
only in the sector of the sfermions of the 3$^{rd}$ generation. 
Therefore, experimental data about stops will give information 
about the soft--breaking trilinear scalar
coupling parameter $A_t$.\\
At an $e^+ e^-$ Linear Collider with center--of--mass energy $\sqrt{s}
\ge 500$~GeV it will be possible to produce the higher stop and sbottom
mass eigenstates $\st_2, \sb_2$. This will allow a detailed study of the 
properties of stops and sbottoms. The decay pattern of $\st_2$ and $\sb_2$
may be rather complicated, because of left--right mixing, and since many
decay channels can be open. In this contribution we will study the decays of
$\st_2$ and $\sb_2$, and we will present numerical predictions for the
important decay branching ratios. We will also give an estimate of the expected
precision for the determination of the underlying soft supersymmetry 
breaking parameters.\\ 

The supersymmetric (SUSY) partners of the Standard Model 
fermions with left and right
helicity are the left and right sfermions. In the case
of stop and sbottom the left and right states are in
general mixed. In the ($\sf_L,\,\sf_R$) basis
the mass matrix is \cite{Ellis,Gunion}
\beqa
M^2_{\tilde{f}} = \left( \begin{array}{cc}
                        m^2_{\tilde{f}_L} & a_f m_f \\
                        a_f m_f & m^2_{\tilde{f}_R}
                       \end{array}
                 \right)
\label{eq:sqmmat}
\eeqa
with
\beqa
  &&\hspace{-8mm} m^2_{\ti f_L} =
  M^2_{\ti Q} + m_Z^2 \cos 2\beta(T^3_f - e_f \sin^2\theta_W) + m^2_f ,
  \label{eq:msfl}\\
  &&\hspace{-8mm} m^2_{\ti f_R} =
  M^2_{\ti F'} + e_f m_Z^2 \cos 2\beta \sin^2\theta_W + m^2_f ,
  \label{eq:msfr}\\
  &&\hspace{-8mm} a_t \equiv A_t - \mu \cot \beta, \hspace{2mm}
  a_b \equiv A_b - \mu \tan \beta\, ,
  \label{eq:offdiag}
\eeqa
where $e_f$ and $T^3_f$ are the charge and the third component
of the weak isospin of the sfermion $\sf$,
$M_{\ti F'} = M_{\ti U},\, M_{\ti D}$
for $\sf_R = \st_R,\, \sb_R$,
respectively, and $m_f$ is the mass of the corresponding fermion.
From renormalization group equations \cite{dreema}
one expects that due to the Yukawa interactions
the soft SUSY breaking masses $M_{\ti Q}$, $M_{\ti U}$, and $M_{\ti D}$
of the $3^{rd}$ generation sfermions are
smaller than those of the $1^{st}$ and $2^{nd}$ generation.
$\st_L$-$\st_R$ mixing is important because
of the large top quark mass \cite{Ellis,Altarelli}. 
For sbottoms left--right mixing can
be important if $\tan \beta \;\grts\; 10$ \cite{Drees,Bartl94}. 
The mass eigenvalues for
the sfermions $\sf = \st,\,\sb$  are
\beq
  m^2_{\ti f_{1,2}} = \onehf \left(m^2_{\ti f_L} + m^2_{\ti f_R}
  \mp \sqrt{(m^2_{\ti f_L} - m^2_{\ti f_R})^2 +
  4m_f^2 a^2_f} \,\right) \label{eq:msf12}
\eeq
where $\st_1$, and $\sb_1$ denote the lighter
eigenstates. The mixing angles $\theta_{\ti f}$ are given by
\beq {\small
  \cth_{\ti f} =
  \frac{- a_f m_f}{\sqrt{(m_{\ti f_L}^2-m_{\ti f_1}^2)^2 + a_f^2 m_f^2}},
  \hspace{5mm}
  \sth_{\ti f} =
  \frac{m_{\ti f_L}^2-m_{\ti f_1}^2}
       {\sqrt{ (m_{\ti f_L}^2-m_{\ti f_1}^2)^2 + a_f^2 m_f^2}}.
} \label{eq:mixangl} 
\eeq

\section{Decays of Stop and Sbottom}

The squarks of the third generation can have the weak decays
($i,j = 1,2$; $k = 1,\ldots,4$)
\beqa
  \st_i       &\ra& t\,\nt_k,    \hspace{4mm} b\,\chp_j\, ,  \label{eq:stdec}\\
  \sb_i       &\ra& b\,\nt_k,    \hspace{4mm} t\,\chm_j\, .  \label{eq:sbdec}
\eeqa
If the strong decays
\beq
  \st_i \ra t\,\sg, \qquad \sb_i \ra b\,\sg
    \label{eq:sqglu}
\eeq
are kinematically allowed then they are the dominant decay modes
of $\st_1$ and $\sb_1$.
Otherwise, the lighter squark mass eigenstates decay mostly according
to \eq{stdec} and \eq{sbdec}.
If $m_{\tilde \chi^0_1} + m_b + m_W < m_{\ti t_1} < 
m_{\tilde \chi^0_1} + m_t$, then the decay $\st_1 \ra b \, W^+ \, \nt_1$
is important \cite{Porod}, 
otherwise the loop decay $\st_1 \ra c\, \nt_1$ can be dominant for 
$m_{\st_1} < m_b + m_{\tilde \chi^\pm_1}$. 
In case of strong left--right mixing the splitting
between the two mass eigenstates may be so large that the
following additional decay modes are present \cite{Bartl94}:
\beqa
 \st_2 &\ra& \st_1\,Z\,, \qquad \sb_1\,W^+\,,
    \label{eq:st2dec}\\
 \sb_2 &\ra& \sb_1\,Z\,, \qquad \st_1\,W^-\,,
    \label{eq:sb2dec}\\
 \st_2 &\ra& \st_1\,h^0 \, (H^0,\, A^0\,), \qquad \sb_1\, H^+\,,
    \\
 \sb_2 &\ra& \st_1\,h^0 \, (H^0,\, A^0\,), \qquad \st_1\, H^-\,.
    \label{eq:sb2decH}
\eeqa

\begin{figure}[h]
\begin{center}
{\setlength{\unitlength}{1mm}
\begin{picture}(160,55)
\put(-3,0){
\begin{picture}(80,55)  
\put(1,0){\mbox{\epsfig{figure=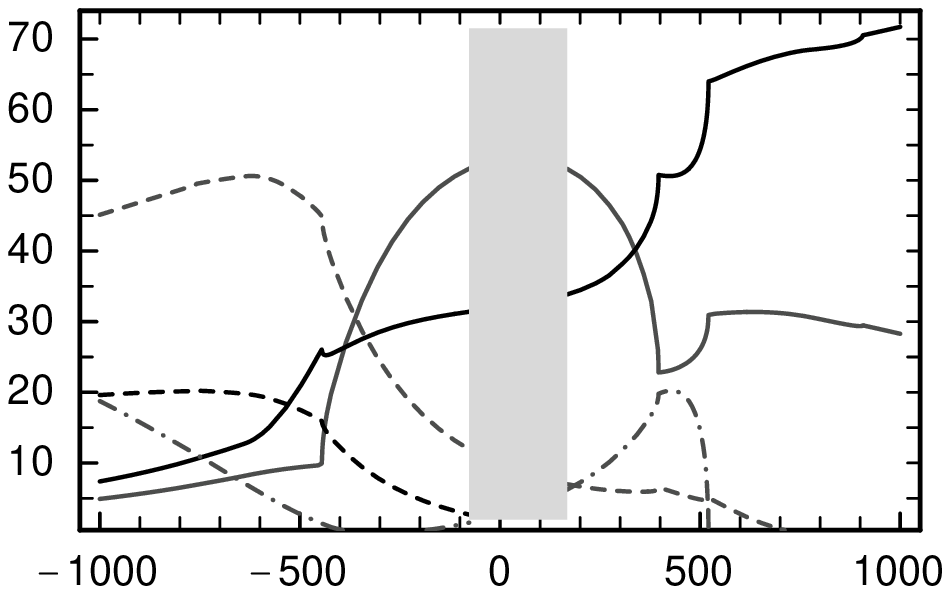,height=4.7cm}}}
\put(3,49.5){\makebox(0,0)[bl]{{\large BR($\st_2$)}~[\%]}}
\put(77,-1){\makebox(0,0)[tr]{{\large $\mu$}~[GeV]}}
\put(71,36){\makebox(0,0)[tr]{{\large (a)}}}
\end{picture}
         }
\put(74,0){
\begin{picture}(80,55)  
\put(1,0){\mbox{\epsfig{figure=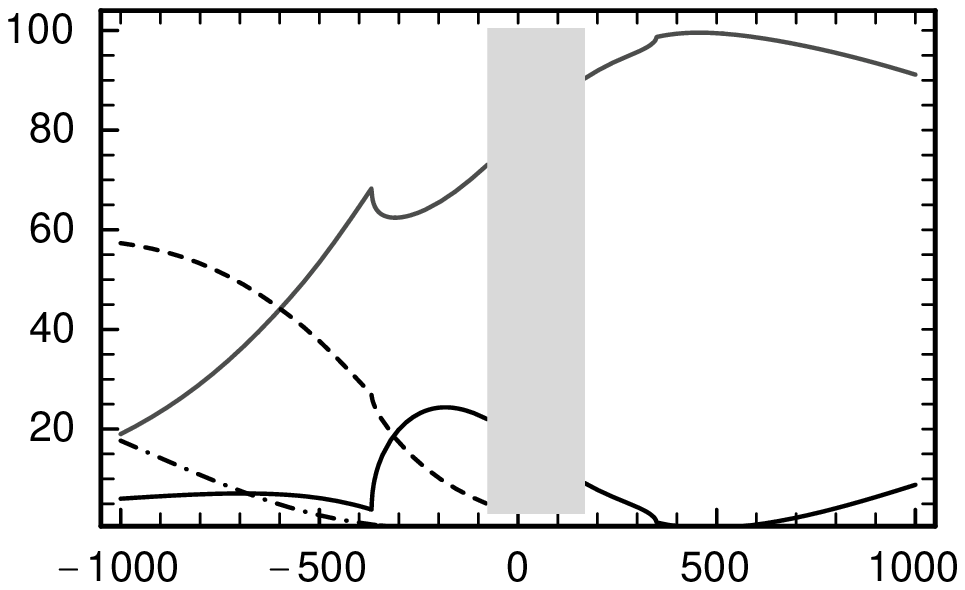,height=4.7cm}}}
\put(3,49.5){\makebox(0,0)[bl]{{\large BR($\sb_2$)}~[\%]}}
\put(77,-1){\makebox(0,0)[tr]{{\large $\mu$}~[GeV]}}
\put(71,36){\makebox(0,0)[tr]{{\large (b)}}}
\end{picture}
         }
\end{picture}}
\end{center}
\caption[fig1]{Branching ratios of (a) $\st_2$ decays and (b) $\sb_2$ decays,
as a function of $\mu$ for $\tan\beta =2$, $M_{\ti Q} = 500$~GeV,
$M_{\ti U} = 444$~GeV, $M_{\ti D} = 556$~GeV, $A_t = A_b = 600$~GeV,
$M = 200$~GeV, $m_{A^0} = 130$~GeV. 
The grey areas are excluded by
LEP data.\\ The curves correspond to the following
transitions:\\
(a) dark full line $\st_2 \to b {\ti \chi}_i^+$ (summed over
$i = 1,2$), light full line $\st_2 \to t {\ti \chi}_k^0$ (summed over
$k = 1,\ldots,4$), dark dashed line $\st_2 \to \sb_i W^+$ (summed over
$i = 1,2$), light dashed line $\st_2 \to \st_1 Z^0$, dash--dotted line
$\st_2 \to \st_1 A^0$,\\
(b) dark full line 
$\sb_2 \to t {\ti \chi}_i^-$ (summed over
$i = 1,2$), light full line $\sb_2 \to b {\ti \chi}_k^0$ (summed over
$k = 1,\ldots,4$), dashed line $\sb_2 \to \st_1 W^-$, dash--dotted line
$\sb_2 \to \st_1 H^-$. 
}
\label{fig:br1ab}
\end{figure}

The decay of the lighter eigenstates $\st_1$ and $\sb_1$ were studied
e.~g. in \cite{Bartl96,Bartl96a}. 
The decay patterns of the heavier squark mass eigenstates can be
quite complicated, because all the decay modes of eqs. \eq{stdec} to
\eq{sb2decH} can occur.
In the following we will study $\st_2$ and $\sb_2$ decays in the
parameter domain where the decays \eq{st2dec} to \eq{sb2decH} are
important. 
Earlier studies are in \cite{Bartl94,Bartl96,Bartl96a,Arhrib,Djouadi}. 
We calculate the different decay widths with the formulae of
Refs.~\cite{Bartl94,Porod,Bartl96}.
We have included radiative corrections to the Higgs masses and the
Higgs mixing angle according to \cite{higgscorr}.
As examples we will plot the branching ratios of $\st_2$
and $\sb_2$ decays as a 
function of the Higgs--higgsino mass parameter $\mu$, for $\tan\beta=2$
and 30,
$A_t = A_b = 600$~GeV, $M_{\ti Q}=500$~GeV, $M_{\ti U}=444$~GeV,
$M_{\ti D}=556$~GeV, taking the SU(2) gaugino mass parameter $M=200$~GeV,
and the mass of the pseudoscalar Higgs boson $m_{A^0} = 130$~GeV. 
For $\tan\beta=2$ the mass
differences between the higher and lower mass eigenstates are
$m_{\st_2} - m_{\st_1} \ge m_Z$ ($m_{\st_2} - m_{\st_1} \ge m_{A^0}$)
for $\mu \le 770$~GeV ($\mu \le 520$~GeV),
$m_{\st_2} - m_{\sb_1} \ge m_W$ ($m_{\st_2} - m_{\sb_1} \ge m_{H^+}$)
for $\mu \le 263$~GeV ($\mu \le -759$~GeV),
$m_{\st_2} - m_{\sb_2} \ge m_W$ for $\mu \le -528$~GeV,
$m_{\sb_2} - m_{\st_1} \ge m_{H^+}$
for $\mu \le 216$~GeV,
$m_{\sb_2} - m_{\st_1} \ge m_W$ in the whole
$\mu$ range considered.
Furthermore, the mass of the light neutral
Higgs boson is in the range 73~GeV $\le m_{h^0} \le 94$~GeV, the mass of
the charged Higgs boson is $m_{H^+} = 153$~GeV.\\
We show in Fig.~\ref{fig:br1ab}a the branching ratios
of $\st_2$ decays for $\tan\beta = 2$. 
As can be seen, for $\mu \le -400$~GeV the decays
into $W^\pm, Z^0$ dominate, because of the large mass difference
$m_{\st_2} - m_{\st_1}$. In the region $-400$~GeV $< \mu < 1000$~GeV the
decays into charginos and neutralinos dominate,
the decays into $W^\pm, Z^0$ being kinematically suppressed because 
$m_{\st_2} - m_{\st_1}$ is too small. For $\mu \lets 600$~GeV the branching
ratio for $\st_2 \to \st_1 A^0$ can go up to 20\%. In Fig.~\ref{fig:br1ab}b we
show the branching ratios for $\sb_2$ decays as a function of $\mu$ and the
same set of parameters as above. For $\mu < -600$~GeV the decay
$\sb_2 \to \st_1 W^-$ dominates, and the branching ratio of $\sb_2 \to
\st_1 H^-$ is of the order of 10\%. For $\mu > -600$~GeV the decay into
neutralinos are the most important ones. The branching ratios for
$\st_2 \to \st_1 h^0, \st_1 H^0, \sb_1 H^+$ and 
$\sb_2 \to \sb_1 h^0, \sb_1 H^0, \st_1 H^-$ are always less than 5\% for
the parameters considered.

\begin{figure}[h]
\begin{center}
{\setlength{\unitlength}{1mm}
\begin{picture}(160,55)
\put(-3,0){
\begin{picture}(80,55)  
\put(1,0){\mbox{\epsfig{figure=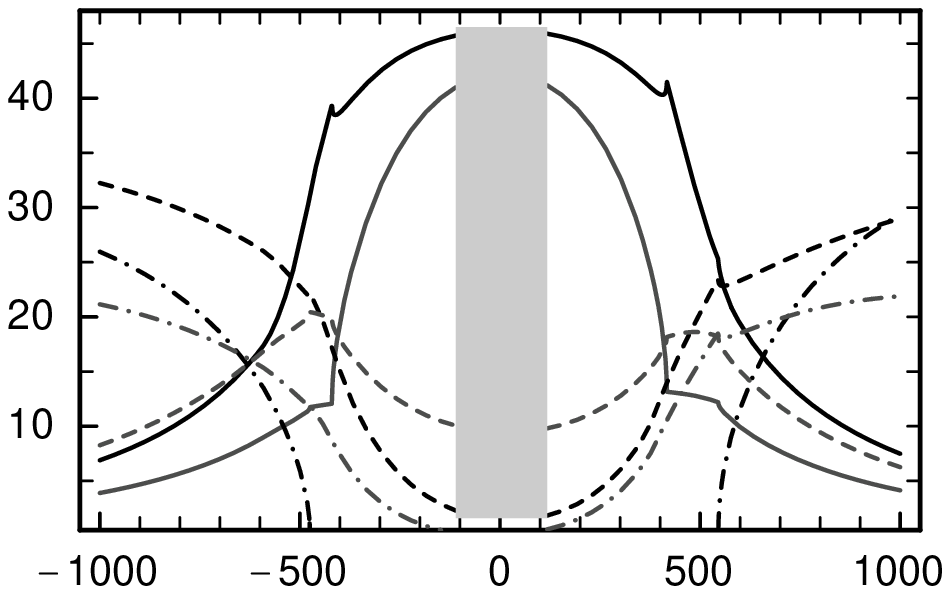,height=4.7cm}}}
\put(3,49.5){\makebox(0,0)[bl]{{\large BR($\st_2$)}~[\%]}}
\put(77,-1){\makebox(0,0)[tr]{{\large $\mu$}~[GeV]}}
\put(71,43){\makebox(0,0)[tr]{{\large (a)}}}
\end{picture}
         }
\put(74,0){
\begin{picture}(80,55)  
\put(1,0){\mbox{\epsfig{figure=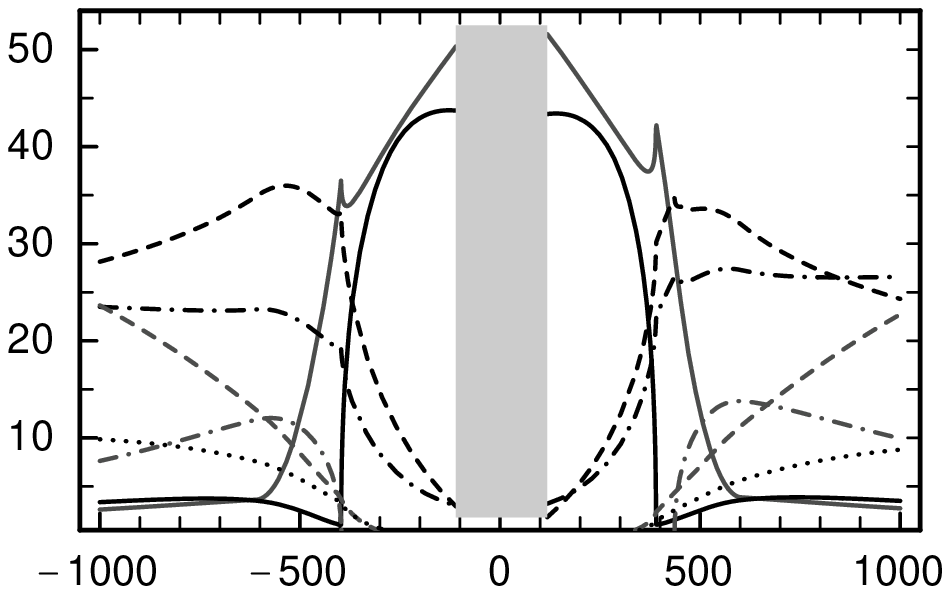,height=4.7cm}}}
\put(3,49.5){\makebox(0,0)[bl]{{\large BR($\sb_2$)}~[\%]}}
\put(77,-1){\makebox(0,0)[tr]{{\large $\mu$}~[GeV]}}
\put(71,43){\makebox(0,0)[tr]{{\large (b)}}}
\end{picture}
         }
\end{picture}}
\end{center}
\caption[fig2]{Branching ratios of (a) $\st_2$ decays and (b) $\sb_2$ decays,
as a function of $\mu$ for $\tan\beta =30$, $M_{\ti Q} = 500$~GeV,
$M_{\ti U} = 444$~GeV, $M_{\ti D} = 556$~GeV, $A_t = A_b = 600$~GeV,
$M = 200$~GeV, $m_{A^0} = 130$~GeV.
The grey areas are excluded by
LEP data.\\ The curves correspond to the following
transitions:\\
(a) dark full line $\st_2 \to b {\ti \chi}_i^+$ (summed over
$i = 1,2$), light full line $\st_2 \to t {\ti \chi}_k^0$ (summed over
$k = 1,\ldots,4$), dark dashed line $\st_2 \to \sb_1 W^+$,
light dashed line $\st_2 \to \st_1 Z^0$, dark dash--dotted line
$\st_2 \to \sb_1 H^+$, light dash--dotted line
$\st_2 \to \st_1 A^0$,\\
(b) dark full line
$\sb_2 \to t {\ti \chi}_i^-$ (summed over
$i = 1,2$), light full line $\sb_2 \to b {\ti \chi}_k^0$ (summed over
$k = 1,\ldots,4$), dark dashed line $\sb_2 \to \st_1 W^-$, 
light dashed line $\sb_2 \to \sb_1 Z^0$,
dark dash--dotted line
$\sb_2 \to \st_1 H^-$, light dash--dotted line $\sb_2 \to \sb_1 A^0$,
dotted line $\sb_2 \to b \sg$.}
\label{fig:br2ab}
\end{figure}

Fig.~\ref{fig:br2ab}a shows the 
branching ratios of $\st_2$ decays as a function of $\mu$ for
$\tan\beta = 30$ and the other parameters as in Figs.~\ref{fig:br1ab}a,b. 
In this case the
decays $\st_2 \to \st_1 Z^0, \st_1 A^0, \sb_1 W^+$, and $\sb_1 \to \st_1 W^+,
\st_1 H^+$ are allowed in the whole $\mu$ range considered.
Moreover, the mass differences are 
$m_{\st_2} - m_{\sb_1} \le m_{H^+}$ for -475 GeV $\le \mu \le 546$~GeV,
$m_{\sb_2} - m_{\sb_1} \le m_{Z}$ for -241 GeV $\le \mu \le 281$~GeV,
$m_{\sb_2} - m_{\sb_1} \le m_{A^0}$ for -396 GeV $\le \mu \le 436$~GeV.
The mass of $h^0$ is in the range 113~GeV $\le m_{h^0} \le 127$~GeV.
The decays into vector bosons and Higgs bosons, eqs.~\eq{st2dec} 
to \eq{sb2decH},
dominate for $|\mu| \grts 550$~GeV, whereas for $|\mu| \lets 550$~GeV
the decays into charginos and neutralinos dominate. The branching ratios
for $\st_2 \to \st_1 h^0, \st_1 H^0$ are always less than 5\%. 
Fig.~\ref{fig:br2ab}b shows the 
branching ratios of $\sb_2$ decays for the same set of
parameters. For $|\mu| \grts 400$~GeV the decays into vector bosons and Higgs
bosons have the largest branching ratios, whereas for 
$|\mu| \lets 400$~GeV the decays into charginos and neutralinos are the 
dominant ones. Note that the decay $\sb_2 \to b \sg$ is possible, but its 
branching ratio is less than 10\%. The reason is that it is phase space 
suppressed.
The branching ratios of $\st_2 \to \st_1 h^0,
\st_1 H^0$, $\sb_2 \to \sb_1 h^0, \sb_1 H^0$ are less than 5\% in the 
parameter space considered.     

\section{Determination of Soft SUSY Breaking Parameters}

We perform a case study of $e^+ e^- \to \st_1 \bar\st_1$ at $\sqrt s =$~500~GeV,
$m_{\st_1} =$~180~GeV, and the left--right stop mixing angle 
$|\cos\tht|$ = 0.57
which corresponds to the minimum of the cross section. The cross
sections at tree level for these parameters are $\sigma_L =$ 48.6~f\/b and
$\sigma_R =$ 46.1~f\/b for 90\% left-- and right--polarized $e^-$~beam,
respectively. Corrections due to initial state radiation,
beamstrahlung, and SUSY--QCD correction have to be applied to the experimental
data. Based on our Monte Carlo studies \cite{Bartl96a}, the experimental
errors on these cross sections are 
$\Delta \sigma_L = \pm$~6~f\/b and $\Delta \sigma_R = \pm$~4.9~f\/b. 
Figure~3 shows
the resulting error bands and the corresponding error ellipse
in the $m_{\st_1}$--$\cos\tht$ plane.
The experimental accuracy for the stop mass and mixing angle is
$m_{\st_1} = 180 \pm 7$~GeV, $|\cos\tht| = 0.57 \pm 0.06$.\\

To treat the sbottom system analogously, we assume that
$\tan\beta$ is low and the $\sb_L$--$\sb_R$ mixing can be neglected, 
$\sb_1 = \sb_L$, $\sb_2 = \sb_R$, i.~e. $\cos\theta_\sb = 1$.
Taking $m_{\sb_1} =$~200~GeV, $m_{\sb_2} =$~220~GeV, the cross sections 
and the expected experimental errors
are $\sigma_L(e^+ e^- \to \sb_1 \bar\sb_1) = 61.1 \pm 6.4$~f\/b,
$\sigma_R(e^+ e^- \to \sb_2 \bar\sb_2) = 6 \pm 2.6$~f\/b
for the 90\% left-- and right--polarized $e^-$ beams.\\

The resulting experimental errors are $m_{\sb_1} = 200 \pm 4$~GeV,
$m_{\sb_2} = 220 \pm 10$~GeV. With these results
we can predict the mass of the
heavier stop, $m_{\st_2} = 289 \pm 15$~GeV. 
This prediction allows experiments to test the MSSM.\\ 
Proceeding further we take $\mu = - 200$~GeV, $\tan\beta = 2$, 
$m_t = 175$~GeV, 
assuming that $\mu$ and $\tan\beta$ are known 
from other experiments.
We obtain the soft breaking para\-meters of the stop and sbottom systems:
$m_{\tilde Q} = 195 \pm 4$~GeV, $m_{\tilde U} = 138 \pm 26$~GeV,
$m_{\tilde D} = 219 \pm 10$~GeV, $A_t = -236 \pm 38$~GeV 
if $\cos\tht > 0$,
and $A_t = 36 \pm 38$~GeV if $\cos\tht < 0$.

\vspace{2mm}

\noindent
{\setlength{\unitlength}{1mm}
\begin{minipage}[t]{7.5cm}
\hspace*{-5mm} \mbox{\psfig{figure=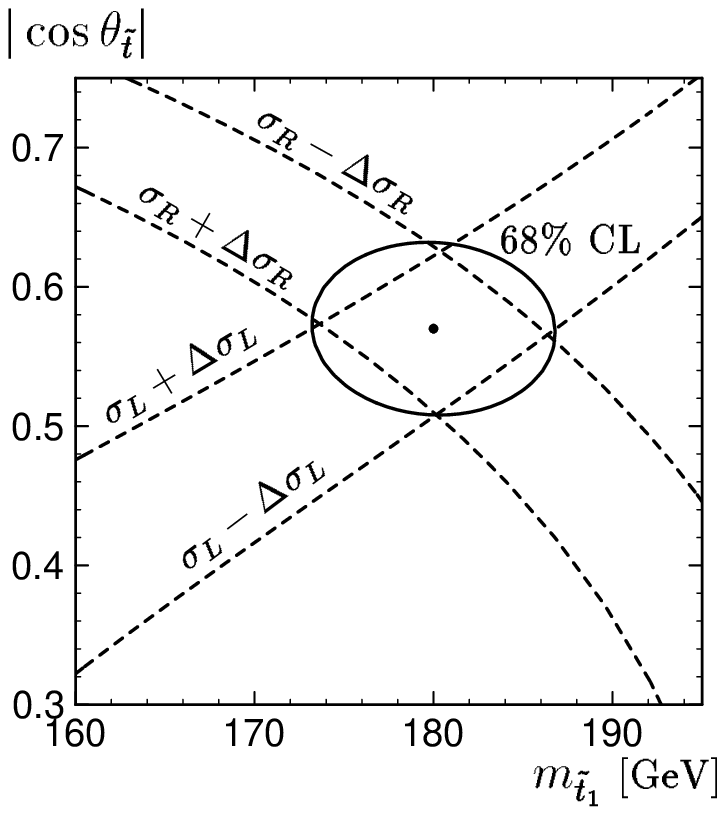}}
\end{minipage} \hfill
\begin{minipage}[b]{6.9cm}
 Figure~3: Error bands (dashed)
 and the corresponding error ellipse
 as a function of $m_{\st_1}$ and $|\cos\theta_{\st}|$ for the
 tree--level cross sections of $e^+ e^- \to \st_1 \st_1$ at
 $\sqrt{s} = 500$~GeV with 90\% left-- and right--polarized electron beam
 The dot corresponds to $m_{\st_1} = 180$~GeV and $|\cos\theta_{\st}| = 0.57$.\\
 The error bands are defined by\\
 $(\sigma_L, \Delta \sigma_L) = (48.6, 6)$~f\/b  and\\
 $(\sigma_R, \Delta \sigma_R) = (46.1, 4.9)$~f\/b.\\
  \vspace{12mm}
\end{minipage}\hspace{3mm}
}

\section*{Acknowledgements}
We are grateful to K.~Hidaka for valuable comments.
This work was supported by the "Fonds zur F\"orderung der
wissenschaftlichen Forschung" of Austria, project no. P10843--PHY.

\end{document}